\documentclass[pra,aps,twocolumn,superscriptaddress,amsmath,amssymb,showpacs]{revtex4}

\usepackage{epsfig}% Include figure files
\usepackage{graphicx}% Include figure files
\usepackage{dcolumn}% Align table columns on decimal point
\usepackage{bm}% bold math

\begin{document}

\title{      Spin-Orbital Liquid on a Triangular Lattice }

\author{     Andrzej M. Ole\'s }
\affiliation{Max-Planck-Institut f\"ur Festk\"orperforschung,
             Heisenbergstrasse 1, D-70569 Stuttgart, Germany }
\affiliation{Marian Smoluchowski Institute of Physics,
             Jagellonian University \\ Reymonta 4, PL-30059 Krak\'ow, Poland }

\author{     Ji\v{r}\'i Chaloupka }
\affiliation{Max-Planck-Institut f\"ur Festk\"orperforschung,
             Heisenbergstrasse 1, D-70569 Stuttgart, Germany }
\affiliation{Department of Condensed Matter Physics, Masaryk
             University \\ Kotl\'a\v{r}sk\'a 2, CZ-61137 Brno,
             Czech Republic }

%\date{\today}

\begin{abstract}
Using Lanczos exact diagonalization of finite clusters we demonstrate
that the spin-orbital $d^1$ model for triply degenerate $t_{2g}$
orbitals on a triangular lattice provides an example of a spin-orbital
liquid ground state. We also show that the spin-orbital liquid involves
entangled valence bond states which violate the Goodenough-Kanamori
rules, and modify effective spin exchange constants.\\
{\it Published in: Acta Phys. Polon. A {\bf 121}, 1026 (2012).}
\end{abstract}


\maketitle

\section{Introduction}

A variety of very interesting and challenging problems in
condensed matter theory arises in systems of strongly correlated
electrons with degenerate orbitals \cite{Ole09}. When the
intraorbital Coulomb element $U$ is much larger than the effective
electron hopping $t$, i.e., $U\gg t$, the magnetic properties
follow from the spin-orbital superexchange \cite{Ole05}. Usually
spin interactions are then determined by orbital order and
complementary types of spin and orbital order coexist in agreement
with the Goodenough-Kanamori rules (GKR) \cite{Goode}. However,
large quantum fluctuations that emerge from strongly frustrated
orbital interactions could instead stabilize disordered phases
\cite{Fei97}. This observation triggered the search for an example
of a spin-orbital liquid (SOL) ground state (GS), similar to a
spin liquid state in a
one-dimensional Heisenberg antiferromagnet. In principle such a
SOL state might be expected for the spin-orbital $d^9$ model on
a triangular lattice in LiNiO$_2$, but Ising-like orbital
interactions suppress it \cite{Rei05}.

Exotic behavior of spin-orbital systems follows from spin-orbital
entanglement (SOE) \cite{Ole06}. To name a few phenomena, it is
responsible for the temperature dependence of optical spectral
weights in LaVO$_3$ \cite{Kha04}, plays a role in the phase
diagram of the $R$VO$_3$ perovskites \cite{Hor08}, and restricts
propagation of a hole in states with entangled spin-orbital order
\cite{Woh09}. Recently novel phases with SOE were discovered in a
bilayer spin-orbital $d^9$ model \cite{Brz11}, but also here a SOL
phase could not be established.

\section{Model and calculation method}

The orbital interactions for $t_{2g}$ orbitals, with $T=1/2$
pseudospins that depend on the bond direction, are more quantum
than for $e_g$ ones and all three pseudospin components contribute
for each bond \cite{Ole09}. Here we employ Lanczos exact
diagonalization to investigate a spin-orbital $d^1$ model for
triply degenerate $t_{2g}$ orbitals on a triangular lattice as in
NaTiO$_2$ \cite{Nor08}, with superexchange (${\cal H}_s$), direct
exchange (${\cal H}_d$) and mixed terms responsible for enhanced
quantum fluctuations (${\cal H}_m$),
\begin{equation}
\label{som} {\cal H} = J \left\{ (1 - \alpha) \; {\cal H}_s
                 + \sqrt{(1 - \alpha) \alpha} \; {\cal H}_m
                 + \alpha \; {\cal H}_d \right\}\,.
\end{equation}
The model depends on the parameter $0\le\alpha\le 1$ and on Hund's
exchange $\eta\equiv J_H/U$. Here $J$ is the exchange energy.
In the direct exchange limit ($\alpha=1$) the model Eq. (1) is exactly
solvable and the GS was determined by considering the valence bond (VB)
dimer coverings of the lattice with each dimer containing a spin
singlet \cite{Jac07}. Below we show by analyzing the results of
Lanczos diagonalization \cite{Cha11} that a SOL is realized in the
opposite superexchange limit ($\alpha=0$) of the model.

The essential feature of the model Eq. (\ref{som}) is that $S=1/2$
quantum spins are coupled by an SU(2) symmetric $({\vec
S}_i\cdot{\vec S}_j)$ interactions, while the orbital
interactions obey a much lower discrete symmetry between three
equivalent directions $\gamma=a,b,c$  in the lattice, with two
$t_{2g}$ orbital flavors active in the superexchange via $({\vec
T}_i\cdot{\vec T}_j)^{(\gamma)}$ term for $T=1/2$ pseudospin, and the
third one contributing to direct exchange; for more details see
\cite{Nor08,Cha11}. The drastic difference in occupied orbital states
realized in both limits at $\eta=0$ is illustrated for the case of a
9-site cluster by insets in Fig.~1(b) --- one finds equal occupancies
of each orbital state at $\alpha=0$, while 756 degenerate orbital dimer
VB-like states are found at $\alpha=1$ \cite{Jac07,Cha11}, and the
degeneracy scales with system size.

We characterize the GS by spin, orbital and spin-orbital (four-operator)
bond correlations ($d$ is the degeneracy of the GS $\{|n\rangle\}$),
given by
\begin{eqnarray}
\label{ss} {\cal S}\!& \equiv & \!\frac{1}{d}\;\sum_n
\big\langle n\big|{\vec S}_i \cdot {\vec S}_j \big|n\big\rangle\,, \\
\label{tt} {\cal T}\! & \equiv & \!\frac{1}{d}\;\sum_n\big\langle
n\big|
({\vec T}_i \cdot {\vec T}_j)^{(\gamma)}\big|n\big\rangle\,,\\
\label{st} {\cal C}\!& \equiv & \frac1d \sum_n \langle n| (\vec
S_i\cdot\vec S_j-{\cal S})(\vec T_i\cdot\vec T_j-{\cal
T})^{(\gamma)} |n\rangle\,.
\end{eqnarray}
Below we consider two clusters with periodic boundary conditions
(PBC): a hexagonal cluster of $N=7$ sites (N7) and a rhombic cluster of
$N=9$ sites (N9). Due to PBC all sites and bonds are equivalent and the
above correlations do not depend on the bond $\langle ij\rangle$
and its direction $\gamma$. Each $t_{2g}$ orbital is occupied on average
by 1/3 electron, but the states are manifestly different in the limits
of $\alpha=0$ and $\alpha=1$, see the insets in Fig. 1(b).

\section{Results and discussion}

In both N7 and N9 cluster spin ${\cal S}$ and orbital ${\cal T}$
correlations are negative and the GKR stating that these
correlations should be complementary are violated, see Fig.~1.
Frustration in the triangular lattice is responsible for a rather
weak and independent of $\alpha$ antiferromagnetic (AF) spin
correlations in the N7 cluster, ${\cal S}\simeq -0.107$, see Fig.~1(a).
These correlations are weaker (${\cal S}\simeq -0.090$) at $\alpha=0$
in the N9 cluster and become more pronounced (${\cal S}\simeq
-0.144$) when $\alpha\simeq 0.6$ and joint spin-orbital
fluctuations weaken to ${\cal C}\simeq -0.050$, see Fig. 1(b). The
orbital correlations weaken as well when $\alpha$ increases toward
$\alpha=1$, particularly in the N9 cluster. Joint spin-orbital
correlations are similar in both clusters (e.g. ${\cal C}\simeq -0.070$
at $\alpha=0$) and $|{\cal C}|$ gradually decreases when spin and
orbitals disentangle approaching $\alpha=1$.

\begin{figure}[t!]
\includegraphics[width=8.2cm]{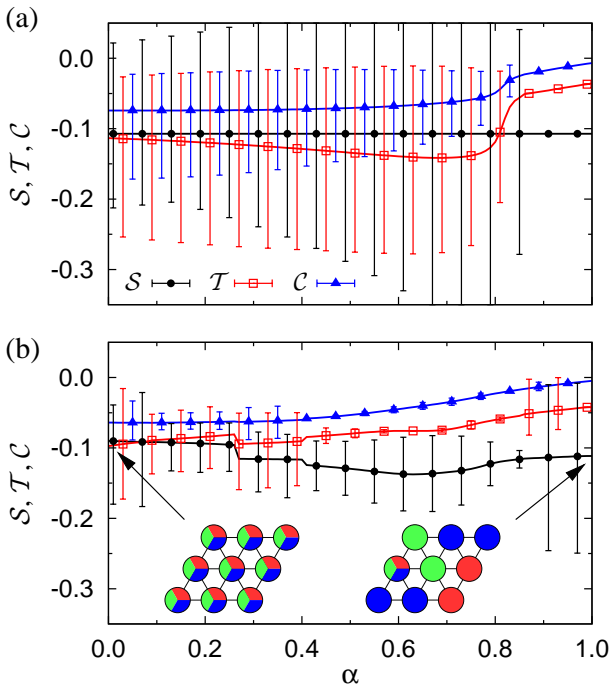}
\caption{Bond correlations at $\eta=0$: spin (${\cal S}$),
orbital (${\cal T}$), and spin-orbital (${\cal C}$) for: (a) N7
cluster, and (b) N9 cluster. The vertical lines indicate the
exactly determined range of possible values that follows from the
GS degeneracy. The insets in (b) indicate typical orbital patterns
in the superexchange ($\alpha=0$) and direct exchange ($\alpha=1$)
limit for the rhombic N9 cluster. } \label{fig:stc}
\end{figure}

An important question is whether spin order and excitations could
be described by an effective spin model derived from Eq. (\ref{som}).
In order to illustrate consequences of SOE in magnetic states we
rewrite the $d^1$ spin-orbital model Eq. (\ref{som}) in a general
form \cite{Ole05} resembling a spin Hamiltonian,
\begin{equation}
\label{somjk}
{\cal H} = \sum_{\langle ij \rangle \parallel\gamma} \left\{
{\hat {\cal J}}_{ij}^{(\gamma)} \left( {\vec S}_i \cdot {\vec S}_j
\right) + {\hat {\cal K}}_{ij}^{(\gamma)} \right\}\,,
\end{equation}
where the orbital operators ${\hat {\cal J}}_{ij}^{(\gamma)}$ and
${\hat {\cal K}}_{ij}^{(\gamma)}$ depend on the parameters
$\{\alpha,\eta\}$ for a bond $\langle ij\rangle$ along axis
$\gamma$. Mean field (MF) procedure used frequently reads \cite{Cha11},
\begin{eqnarray}
\label{sommf}
{\cal H}_{\rm MF} &=& \sum_{\langle ij\rangle \parallel\gamma}\left\{
 \left\langle{\hat {\cal J}}_{ij}^{(\gamma)}\right\rangle
             {\vec S}_i\cdot {\vec S}_j
-\left\langle{\hat {\cal J}}_{ij}^{(\gamma)}\right\rangle
\left\langle {\vec S}_i\cdot {\vec S}_j\right\rangle
\right\}\nonumber \\
&+&                  \sum_{\langle ij\rangle \parallel\gamma} \left\{
{\hat {\cal J}}_{ij}^{(\gamma)} \left\langle {\vec S}_i\cdot {\vec S}_j
\right\rangle + {\hat {\cal K}}_{ij}^{(\gamma)} \right\}\,.
\end{eqnarray}
It disentangles spin and orbital degrees of freedom and is used to
determine the MF spin constant for N7 and N9 clusters by averaging the
orbital operator ${\hat {\cal J}}_{ij}^{(\gamma)}$ (its explicit form
is given in \cite{Cha11}) over the MF GS $|\Phi_0\rangle$,
\begin{equation}
\label{jmf}
J_{\rm MF}\equiv
\langle\Phi_0|{\hat{\cal J}}_{ij}^{(\gamma)}|\Phi_0\rangle\,.
\end{equation}
Note that the orbital fluctuations in the term
$\propto\sqrt{\alpha(1-\alpha)}$ in Eq. (1) contribute here as
well as they couple different components of $|\Phi_0\rangle$. In
contrast, the exact exchange constant, $J_{\rm exact}$, is found
when the exact GS $|\Phi\rangle$ obtained after Lanczos
diagonalization is used.

\begin{figure}[t!]
\includegraphics[width=7.5cm]{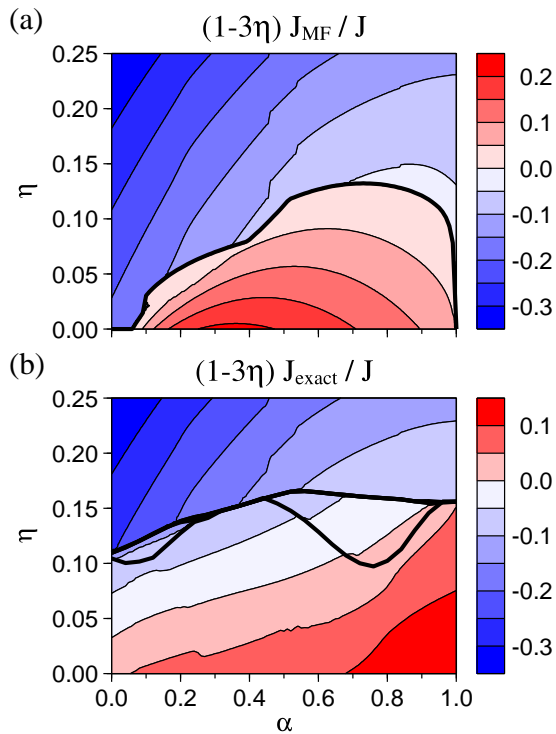}
\caption{Phase diagram in the $(\alpha,\eta)$ plane (heavy lines)
and exchange constants (contour plots) as obtained for the N9
cluster with PBC: (a) within the MF calculation, and (b) using exact
Lanczos diagonalization. In the MF case (a) the transition from
low-spin (${\cal S}_t=1/2$) to high-spin (${\cal S}_t=9/2$) phase
occurs when $J_{\rm MF}$ changes sign. In the exact calculation
one finds in addition an intermediate phase with ${\cal S}_t=3/2$,
stable between the heavy black lines in (b).} \label{fig:jeff}
\end{figure}

In Fig. 2 we compare the phase diagrams obtained from the above MF
procedure and from exact diagonalization for the N9 cluster. Consider
first a quantum phase transition from the low-spin (${\cal S}_t=1/2$)
disordered phase to the high-spin (${\cal S}_t=9/2$) ferromagnetic
(FM) phase which occurs for sufficiently large $\eta$.
When spin and orbital operators are disentangled in Eq. (\ref{sommf}),
i.e., ${\cal C}\equiv 0$ \cite{Ole06}, it coincides with the sign
change of the MF exchange constant $J_{\rm MF}$ and no other phase
(with $1/2<{\cal S}_t<9/2$) is found, see Fig. 2(a), as in a spin system.

Comparing the values of $J_{\rm MF}$ and $J_{\rm exact}$ found from the
MF and from exact diagonalization of the N9 cluster (Fig. 2), one finds
that $J_{\rm exact}\ge J_{\rm MF}$ in a broad regime of $\alpha$ except
for $\alpha\simeq 1$. Therefore, the MF approximation turns out to be
rather unrealistic and overestimates (underestimates) the stability of
states with FM (AF) spin correlations.
The value of $J_{\rm MF}$ decreases with increasing $\eta$,
but positive values $J_{\rm MF}>0$ are found at $\eta=0$ only if
$0.07<\alpha<1$. This demonstrates that FM states:
($i$) are favored when joint spin-orbital fluctuations are suppressed,
and
($ii$) are stabilized by orbital fluctuations close to $\alpha=0$ even
in absence of Hund's exchange.
The transition from the low-spin
(${\cal S}_t=1/2$) to the high-spin (${\cal S}_t=9/2$) state occurs in
presence of SOE at a much higher value of $\eta\approx 0.14$, with only
weak dependence on $\alpha$, see Fig. 2b. In addition, one finds a phase 
with an intermediate spin value ${\cal S}_t=3/2$ for $0<\alpha<0.21$ and
$0.44<\alpha<0.88$, and $J_{\rm exact}$ changes discontinuously at the
transition to the FM phase.

Altogether, the qualitative trends found for the N9 cluster are
generic and agree with those observed for the N7 cluster, see Fig.
15 in \cite{Cha11}. In both cases one finds that:
($i$) the FM phase is stable in the MF approximation close to $\alpha=0$
and becomes degenerate with the low-spin phase at $\alpha=1$,
($ii$) the MF procedure is exact in the regime of FM phase, and
($iii$) the transition to the FM phase occurs gradually through
intermediate values of total spin ${\cal S}_t$ (except at $\alpha=1$).
This suggests that partially polarized FM phase should occur in the
thermodynamic limit of the model Eq. (1) and arises due to SOE which
is gradually suppressed when $\eta$ increases.

We argue that the present study and the results presented in
\cite{Cha11} provide evidence in favor of a quantum SOL phase in
the present $d^1$ spin-orbital model Eq. (1) in a corner of its
phase diagram --- in the regime of small values of both $\alpha$ and
$\eta$ parameter. In agreement with the directional nature of orbital
interactions, this SOL phase is also characterized by rather strong
VB dimer correlations \cite{Nor11}. The consequences of SOE
are more severe in this phase and the transition
to the FM phase occurs typically at a {\it much higher value\/} of
$\eta$ than the one where $J_{\rm exact}$ changes its sign.
Therefore, we suggest that even in case when magnetic
exchange $J_{\rm exact}$ is accurately evaluated using
the relevant orbital correlations, it looses its predictive
power and is inadequate to describe the magnetic ground state and
excitations in a system dominated by SOE, where the GKR are violated.
Note that the frustrated triangular lattice plays here an
important role and removes any kind of orbital order.

\section{Summary and conclusions}

Summarizing, we have demonstrated that the GKR are violated in the
regime of weak Hund's exchange and the {\it spin-orbital liquid\/}
phase is stabilized by spin-orbital entanglement in the $d^1$
spin-orbital model on the triangular lattice. In this regime the MF
decoupling procedure of spin and orbital operators fails and the
magnetic properties can be determined only by solving
the full entangled spin-orbital many-body problem.

Finally, we point out that spin-orbital entangled states play a
role in exotic ground states \cite{Brz11} as well as low energy
excitations for realistic values of $\eta\simeq 0.14$. Their
consequences have already been established in the vanadium perovskites
\cite{Kha04,Hor08,Woh09}, and we expect that they could be of even
more importance in systems with geometrically frustrated lattice
such as the one considered here.

\section*{Acknowledgments}

We thank Bruce Normand for insightful discussions.
A.M. Ole\'s acknowledges support by the Foundation for Polish Science
(FNP) and by the Polish National Science Center (NCN) Project No. N202~069639.
J.~Chaloupka acknowledges the fellowship of the Alexander von Humboldt
Foundation and support by the Ministry of Education of Czech Republic
under Grant No. MSM0021622410.

\end{document}